\documentclass{PoS}
\usepackage{bbm}
\usepackage{mathrsfs}
\usepackage{amsmath}

\newcommand{\tr}{\mathrm{tr}}
\newcommand{\Vek}[1]{\boldsymbol{#1}}
\newcommand{\vek}[1]{\mathbf{#1}}

\title{Temporal Wilson loop in the Hamiltonian approach in Coulomb gauge}

\ShortTitle{Temporal Wilson loop in the Hamiltonian approach in Coulomb gauge}

\author{\speaker{Markus Quandt}
\\
        Universit\"at T\"ubingen\\
        E-mail: \email{markus.quandt@uni-tuebingen.de}}

\author{H.~Reinhardt\\
        Universit\"at T\"ubingen\\
        E-mail: \email{hugo.reinhardt@uni-tuebingen.de}}

\author{G.~Burgio\\
        Universit\"at T\"ubingen\\
        E-mail: \email{giuseppe.burgio@uni-tuebingen.de}}

\abstract{We investigate the temporal Wilson loop using the Hamiltonian approach 
to Yang-Mills theory. In simple cases such as the Abelian theory  or 
the non-Abelian theory in (1+1) dimensions, the known results can be 
derived using unitary transformations to take care of time evolution. 
Alternatively, the exact solution can also be found in Coulomb gauge 
using the exact ground state wave functional which is known explicitly 
in these simple cases. The Coulomb gauge technique can also be applied 
to the more realistic case of Yang-Mills theory in (3+1) dimensions, 
where one has to rely on the approximate vacuum wave functional obtained 
e.g. in recent variational approaches. We use this formulation to compute 
the temporal Wilson loop in (3+1) dimensional Yang-Mills theory, and find 
that the Wilson and Coulomb string tension agree within this approximation 
scheme. Possible improvements of these findings are briefly discussed.}

\FullConference{Xth Quark Confinement and the Hadron Spectrum\\
                 8-12 October 2012\\
                 TUM Campus Garching, Munich, Germany}

\begin{document}

\section{Introduction}
\label{sec:intro}
Recent efforts to study Yang-Mills (YM) theory 
are largely based on Coulomb gauge, 
because the Gribov-Zwanziger picture of confinement \cite{Gribov:1977wm, Zwanziger:1995cv}
becomes particularly transparent in this framework: 
the only constraint on the physical states in a Hamiltonian approach based on Coulomb gauge 
is \emph{Gau\ss{}' law}, which in turn can be resolved exactly so that any normalizable 
wave functional can be used as a physical state. 

Such a formulation naturally lends itself to variational methods, which are usually based on 
(modified) Gaussian \emph{Ans\"atze} for the vacuum wave functional, see 
e.g.~\cite{Szczepaniak:2001rg,Epple:2006hv} and references therein. 
The solutions agree with the Gribov-Zwanziger scenario and, when combined with the horizon condition, 
lead to a coherent picture of strongly coupled YM theory, which is mainly dominated by 
a diverging gluon energy in the infrared, a linearly rising Coulomb potential between static quarks,
and a strongly enhanced Faddeev-Popov ghost propagator in the infrared. The latter condition 
implies that the YM vacuum behaves  as a perfect colour dia-electric 
medium \cite{Reinhardt:2008ek}, i.e.~a \emph{dual superconductor}.
All these findings agree qualitatively with the results from recent lattice simulations
\cite{Burgio:2008jr}, although there are still some discrepancies in the 
quantitative details, in particular for the ghost sector \cite{Burgio:2012bk}.  

However, this simple picture is subject to the caveat that the Coulomb potential
is only an \emph{upper bound} for the true potential between heavy quarks \cite{Zwanziger:2002sh}. 
For a complete description, one has to address 
the physical potential from large Wilson loops directly. In the continuum, this is complicated 
by path ordering and large self-energy contributions which obscure the extraction of the 
true potential from large Wilson loops \cite{Haagensen:1997pi}. 

In this talk, I report on a recent approach \cite{Reinhardt:2011fq} to compute the 
Wilson potential in the Hamiltonian approach in Coulomb gauge. A careful analysis of Gau\ss{}' law
and the physical degrees of freedom allows for an elegant way to isolate and remove the 
self-energy contributions to the Wilson loop, while unitary transformations are applied to take 
care of time evolution. As specific examples, we study quantum electrodynamics 
(or rather Maxwell's theory) in $D=(3+1)$ spacetime dimensions,
and YM theory in both $D=(1+1)$ and $D=(3+1)$. 
In the latter case, only an approximate solution can be found and I briefly discuss possible 
ways to improve our findings.

\section{The Wilson loop in the Hamiltonian approach}
\label{sec:general}
We study $SU(N)$ YM theory in euclidean space using the Hamiltonian approach in 
Weyl gauge, $A_0 = 0$. The physical energy states of this formulation are 
stationary wave functionals $\Psi_n[\vek{A}(\vek{x})]\cdot e^{- iE_n t}$ which obey the 
Schr\"odinger equation based on the  Hamiltonian
\begin{equation}
H = \frac{g^2}{2}\,\int dx\,\Vek{\Pi}^2(\vek{x}) + \frac{1}{2\,g^2}\,\int dx\,
\vek{B}^2 \,.
\label{1}
\end{equation}
Here, $\vek{A}^a(\vek{x})$ is the gauge connection, $\Vek{\Pi}^a(\vek{x}) = - i \delta / \delta\vek{A}^a(\vek{x})$ 
its conjugate momentum, $g$ the (bare) coupling constant and 
$\vek{B}^a = \nabla \times \vek{A}^a - \frac{1}{2} f^{abc}\vek{A}^b \times \vek{A}^c$ 
the non-Abelian magnetic field. 
    
To proceed, we must fix the residual time-independent gauge symmetry left after imposing Weyl 
gauge, for which we choose the Coulomb condition, $\nabla \cdot \vek{A} = 0$. This adds gauge-fixing 
terms to the Hamiltonian (which we discuss below) and imposes \emph{Gau\ss{}' law} as a constraint 
on the physical states:
In the sector with $n$ static charges located at positions $\vek{x}_1\,\ldots,\vek{x}_n$, we have 
$n$ additional colour indices on the wave functionals and Gau\ss{}' law reads (with generators in 
the right representation)
\begin{equation}
\hat{\Gamma}^a(\vek{z}) \,\Psi_{i_1,\ldots i_n}(\vek{A}\,;\,\vek{x}_1,\ldots, \vek{x}_n)
= - \left\{ \sum_{\ell=0}^n \sum_{\{k_\ell\}} T^a_{i_\ell k_\ell}\,
\delta(\vek{z}-\vek{x}_\ell)
\prod_{m \neq \ell}\delta_{i_m k_m}
\right\} \cdot
\,\Psi_{k_1,\ldots,k_n}(\vek{A}\,;\,\vek{x}_1,\ldots \vek{x}_n)\,.
\label{2}
\end{equation}
The complete physical Hilbert space thus decomposes into orthogonal charge sectors 
characterised by the data from the external charges.

Next we consider a rectangular temporal Wilson loop of extension $(T \times R)$. 
Since $A_0 = 0$, we have 
$W = \tr\big\{ U[\vek{A}](\vek{R},\vek{0};T)^\dagger \cdot U[\vek{A}](\vek{0},\vek{0};0)\big\}$,
where $U[\vek{A}](\vek{y},\vek{x};t)$ is the parallel transporter $\vek{x} \to \vek{y}$ at 
fixed time $t$, in the background of the gauge potential $\vek{A}$. Using euclidean time
evolution and the abbreviation $U(\vek{x}) \equiv U[\vek{A}](\vek{x},\vek{0};0)$, we find 
\begin{equation}
\langle W \rangle = \tr\,\langle 0 \,| \, U(\vek{R})^\dagger\,e^{- (H-E_0) T}\,U(\vek{R})\,| \,0 \rangle
= \tr\,\langle \vek{R}\,|\,e^{- T(H-E_0)}\,|\,\vek{R}\rangle\,. 
\label{3}
\end{equation}
The salient point is now that the vacuum is gauge-invariant, while the \emph{Wilson state} 
$|\vek{R}\rangle = U(\vek{R})\,|0\rangle$ is in the $q\bar{q}$-sector according to eq.~(\ref{2}).
Inserting a complete set of energy eigenstates from all sectors, only the $q\bar{q}$-sector will
thus contribute and we have
\begin{equation}
\langle W \rangle  = \tr\, \sum_n\langle\, 0 \,|\, U^\dagger(\vek{R})\,
|\,n\,\rangle_{q\bar{q}}\cdot e^{- (H - E_0) T} \cdot {}_{q\bar{q}}\langle\,n\,|
\,U(\vek{R}) \,|\, 0\, \rangle =
\sum_{n}\,e^{- T (E_n^{(q\bar{q})} - E_0)}\,
\| \langle \vek{R}\,|\,n\,\rangle_{q\bar{q}}
\|^2\,, 
\label{4}
\end{equation} 
where $\| \Omega\|^2 \equiv \tr(\Omega^\dagger\Omega)$. In the limit of large
Euclidean time extensions, we project out the Wilson potential as the difference 
between the ground state energies with and without static quarks,
$
E_0^{q\bar{q}}(R) - E_0 = -\lim_{T \to \infty}\,T^{-1}\,\ln\,\langle W \rangle\,. 
$

The next step is to remove the Wilson line from the state $|\vek{R}\rangle$ by means of a 
\emph{unitary transformation} based on the parallel transporter $U(\vek{R})$. Under such a 
transformation, the conjugate momentum (and thus the Hamiltonian eq.~(\ref{1})) 
acquires an additional term,
$\Pi^a_i(\vek{x}) \to \widetilde{\Pi}^a_i(\vek{x}) = \Pi^a_i(\vek{x}) + \color{black}{\epsilon^a_i(\vek{x})}$, 
where
\begin{align}
  \Vek{\epsilon}^a(\vek{x})\equiv \int_0^{\vek{R}} d\vek{y}\,\delta(\vek{x}-\vek{y})\, U(\vek{y})^\dagger
  i T^a\,U(\vek{y}) \,.
  \label{6} 
\end{align}
As a consequence, the Wilson loop is now computable as a zero-charge 
vacuum expectation value,
\begin{align}
 \langle W \rangle &= \tr\,\langle \,0\,|\,e^{- T (\widetilde{H} - E_0)}\,|\,0\,\rangle
 \label{7} 
\end{align}
where $\widetilde{H}$ differs from eq.~(\ref{1}) by the momentum shift\footnote{This shift complicates 
the treatment of $\widetilde{H}$ considerably, since $[\Vek{\Pi}^a(\vek{x}), \Vek{\epsilon}^b(\vek{x})] \neq 0$.}
$\Vek{\epsilon}(\vek{x})$. 

To illustrate the problems in extracting the true potential, let us consider the contribution to the 
Wilson loop eq.~(\ref{7}) which is quadratic in the induced electric field $\Vek{\epsilon}(\vek{x})$, 
\begin{align}
   V_{\rm ind}(R) = \frac{g^2}{2}\,\tr\,\int dx\,[\Vek{\epsilon}^a(\vek{x})]^2 
   = \frac{g^2}{2}\,C_2 \,\delta_\perp^{(2)}(0)\cdot|\vek{R}|\,,
\label{9}
\end{align}
where $C_2$ is the quadratic Casimir operator for the colour group, and the (quadratically) divergent 
prefactor $\delta_\perp(0)$ is due to the infinitely thin Wilson lines. 
Although $V_{\rm ind}$ is formally confining, it must be spurious because it exists even 
for $G=U(1)$. On the other hand, it is also not correct to simply drop $V_{\rm ind}$, 
because the true potential is hidden underneath the divergence: For $G=U(1)$,
\begin{align}
V_{\rm ind}(R) = \frac{g^2}{2}\,\tr\,\int dx\,\left[ \Vek{\epsilon}_\perp(\vek{x}) \right]^2 + 
\frac{g^2}{2}\,\tr\,\int dx\,\left[ \Vek{\epsilon}_{\|}(\vek{x}) \right]^2\,,
\label{10}
\end{align}
where the first factor renormalizes the Wilson loop, and the second factor gives the true (Coulomb) 
potential, cf.~section \ref{sec:qed}. 

The unitary transformation leading to eq.~(\ref{7}) only assumes Weyl gauge  and thus works with 
or without Coulomb gauge fixing. In cases where the exact gauge-invariant ground state \emph{is}
known, it is not necessary to fix the residual gauge freedom since the Wilson loop is gauge invariant.
In more realistic cases, however, the gauge invariant ground state will not be known and we have 
to resort to approximations based on Coulomb gauge. The required 
gauge-fixing is virtually impossible to effect directly in eq.~(\ref{7}), except for the Abelian model.
A much better strategy is to go back to eq.~(\ref{3}), perform the standard gauge fixing 
and resolve Gau\ss{}' law exactly, which induces in $H \to H_{\rm fix}$ the well-known Coulomb 
term,\footnote{As a consequence of the parallel transport, 
the non-Abelian charges 
$\rho^a(\vek{x}) = \delta(\vek{x} - \vek{R})\, iT^a - 
\delta(\vek{x})\, U(\vek{R})\, i T^a \,U(\vek{R})^\dagger$
are field-dependent.}
\begin{align}
  H_C = \frac{g^2}{2}\,\int d^3(x,y)\,\mathscr{J}^{-1}\,[\rho^a(\vek{x}) + \rho^a_{\rm dyn}(\vek{x})]
  \,\mathscr{J}\,F^{ab}(\vek{x},\vek{y})\,[\rho^b(\vek{y}) + {\rho^b_{\rm dyn}(\vek{y})}]\,,
\label{11}
\end{align}
where $\mathscr{J}$ is the Faddeev-Popov determinant and 
$F^{ab}$ the the so-called Coulomb kernel. 
The Wilson loop now reads
\begin{equation}
\langle W \rangle = \mathrm{tr}\,\langle \,\vek{R}_\perp\,|\, e^{- T (H_{\rm fix} - E_0)}\, | 
\,\vek{R}_\perp\,\rangle 
\label{12}
\end{equation}
where the  new Wilson-like state $|\vek{R}_\perp\rangle = U[\vek{A}_\perp](\vek{R}) | 0 \rangle$ is no 
longer from the $q\bar{q}$ sector, but rather has overlap with the zero-charge vacuum. Any divergences 
associated with the point-like cross section of of the Wilson lines in $|\vek{R}_\perp\rangle$ 
must therefore be unrelated to external charges: as we will see, the extra divergences renormalize the 
composite operator $W\to W_R \equiv Z_W(\Lambda)\cdot W$, but they do not affect the physical potential. 
Whether or not the state  
$|\vek{R}_\perp\rangle$ in eq.~(\ref{12}) is subsequently removed from the Wilson loop by a unitary 
transformation $U_\perp(\vek{R})$ as described earlier becomes a matter of computational convenience. 

\section{Applications}
\label{sec:app}

\subsection{Quantum Electrodynamics}
\label{sec:qed}
In the Abelian case $G=U(1)$, the exact ground state is known and we can first attempt a 
gauge-invariant treatment based on eq.~(\ref{7}). The induced electric field
\begin{align}
\Vek{\epsilon}(\vek{x}) = \int\limits_{\vek{0}}^{\vek{R}}
d\vek{y}\,\delta^{(3)}(\vek{x}-\vek{y})\,.
\label{x1}
\end{align}
is now field-independent, so that the induced potential (\ref{9}) could be pulled out of the expectation 
value $\langle W \rangle$. In view of the discussion above, it is more convenient to first split 
$\Vek{\epsilon} = \Vek{\epsilon}^\perp + \Vek{\epsilon}^{\|}$ and pull out the longitudinal piece only,
$
\langle W \rangle = e^{- T V_{\|}(R)} \cdot \langle 0 | e^{- T 
(\widetilde{H}_{\rm QED}' - E_0)} | 0 \rangle
$
where 
\begin{align}
\widetilde{H}_{\rm QED}' = \frac{1}{2} \int d^3 x \left\{ g^2\,\big[
\Vek{\Pi}^\perp (\vek{x}) + \Vek{\epsilon}^\perp (\vek{x}) \big]^2
+ \frac{1}{g^2}\,\big[\nabla \times \vek{A}^\perp(\vek{x})\big]^2
\right\}\,.
\label{x3}
\end{align}
It is now convenient to reverse the unitary transformation that led to eq.~(\ref{9}), but this time 
with the \emph{transversal} gauge connection $\vek{A}^\perp$ only. As a result, the shift in the electric field
is removed from the Hamiltonian (\ref{x3}) and reshuffled into the state 
$|\vek{R}^\perp\rangle = U[\vek{A}^\perp](\vek{R})\,|0\rangle \equiv U^\perp(\vek{R})\,|0\rangle$. 
Although this new state resembles the initial Wilson state, it is gauge-invariant as it depends on 
$\vek{A}^\perp$ only. Inserting a 
complete set of eigenstates of the standard QED Hamiltonian eq.~(\ref{x3}) (without $\Vek{\epsilon}^\perp$) 
yields 
\begin{align}   
 \langle W \rangle = e^{ - T V_{\|}(R)}&\cdot\sum_n | \langle 0 \,| \, U^{\perp}(\vek{R})\,| \,n \,\rangle | ^2 \,
 e^{- T (E_n - E_0)} \stackrel{T \to \infty}{\longrightarrow} | \langle 0 | U^\perp(\vek{R}) | 0 \rangle |^2
 \cdot e^{- T V_{\|}(R)}\,,
\label{x4}
\end{align}
provided that $|\vek{R}^\perp\rangle$ has non-vanishing overlap with the true vacuum.
For Maxwell's theory the ground state is Gaussian and the relevant matrix element can be computed 
exactly. Using a $O(3)$-invariant UV cutoff $\Lambda$ for the momentum integration, we find 
that $Z_W$ formally vanishes in the limit $\Lambda \to \infty$ because $|\vek{R}^\perp\rangle$ has poor 
overlap with the true ground state \cite{Haagensen:1997pi}. On the other hand, $Z_W$ is also independent 
of the temporal extension $T$ of the loop and hence cannot contribute to the physical potential. 

The correct interpretation  is given by the operator product expansion (OPE): Since the loop operator $W$ 
contains products of field operators at arbitrarily close points, we expect short distance 
($\Lambda \to \infty$) divergences associated with $W$, on top of the counter terms necessary 
to render Green functions finite. 
In fact, it has been known for a long time \cite{Gervais:1979fv, Polyakov:1980ca} that one
overall multiplication $W \to W_R \equiv Z_W^{-1}\cdot W$ is sufficient to render 
$\langle W_R \rangle = e^{- T V_\perp}$ finite. From eq.~(\ref{x1}),  the true physical potential 
becomes
\begin{align}
 V(R) &= V_{\|}(R) = \frac{g^2}{2} \int dx 
\,[ \Vek{\epsilon}^{\|}(\vek{x})]^2 = \frac{g^2}{4 \pi R} - \frac{g^2}{4 \pi |\vek{0}|}\,.
\label{x6}
\end{align}
This is just the usual Coulomb potential including the self-energy of the charges. 

The entire derivation could also be repeated in a Coulomb gauge fixed formulation. If we start from
eq.~(\ref{7}), the gauge fixing gives $\widetilde{H} \to \widetilde{H}_{\rm fix} = H_{\rm QED}' + V_{\|}$,
cf.~eq.~(\ref{x3}), and the computation becomes identical to the gauge-invariant treatment above. 
Alternatively, we could also start from eq.~(\ref{3}) and resolve Gau\ss{}' law for the state $|\vek{R}\rangle$ 
explicitly, which induces the Coulomb term $ H \to H_{\rm fix} = H_{\rm QED}[\vek{A}_\perp] + H_C$ in the Hamiltonian.
This time, the Coulomb potential comes directly from $H_C$, while the first term in $H_{\rm fix}$ only 
renormalizes the Wilson loop.

\subsection{Yang-Mills Theory in $D=1+1$}
\label{sec:ym11}
This model is (almost) topological and requires a non-contractible spactime manifold to give 
nontrivial results. Since time must be unrestricted in the Hamiltonian formalism, we 
compactify space to an interval $[0,L]$ with periodic boundary conditions and take 
spacetime to be the cylinder $\mathsf{M} = \mathbbm{R} \times S^1$. 
For this model, the induced potential eq.~(\ref{9}) gives the correct string tension
\begin{align}
\sigma_{1+1} = \frac{g^2}{2}\,C_2 \stackrel{SU(2)}{=} \frac{3}{8}\,g^2\,.
\label{x7} 
\end{align}
In view of the above discussion, it is, however, unclear \emph{why} this should be 
the correct answer, in particular since $V_{\rm ind} = \sigma_{1+1}\,R$ is not 
periodic or invariant under $R \to L-R$ as expected from the compactification. 

To address this issue, we go back to eq.~(\ref{3}) and fix the minimal Coulomb gauge 
$\partial_1 A_1^a = 0$ and $A_1^a = \text{diag}$. Taking the colour group $G=SU(2)$ for 
simplicity, we are thus left with only one scalar degree of freedom
\begin{align}
\vartheta \equiv \frac{1}{2}\,A_1^3\,L \in [0, \pi]\,. 
\label{x8}
\end{align}
The restriction to the compact interval (the fundamental modular region in this case)
eliminates all residual gauge symmetries. In order to fix eq.~(\ref{3}), we have 
to determine the gauge fixed Hamiltonian in the $q\bar{q}$-sector to which the 
Wilson state $|R\rangle$ belongs. This has two pieces
\begin{align}
H_{\rm fix} = H_{\rm fix}^0 + H_C = H_{\rm fix} +  \frac{g^2}{2} \int d(x,y)\,\rho^a(x)\,
F^{ab}(x,y)\,\rho^b(y)\,,
\label{x9}
\end{align}
where the non-Abelian charge interaction comes from the resolution of Gau\ss{}' law. 
The zero-charge Hamiltonian $H_{\rm fix}^0$ and its spectrum have been studied 
thoroughly in ref.~\cite{Reinhardt:2008ij}. As for the Coulomb term, we have to take 
into account that the external charges $\rho^a$ are field-dependent due to the 
parallel transport. Expanding all colour vectors in a a polar basis ($\vek{e}_0 = e_\vek{3}$, 
$\vek{e}_\pm = (\vek{e}_1 \pm \vek{e}_2) / \sqrt{2}$), we find  
\begin{align}
  H_C =
 \frac{g^2}{8}\sum_{\sigma=-1}^1\tau_\sigma\,\tau_\sigma^\dagger 
   \,\left\{ 2 F_\sigma(0;\vartheta)- 
  \left[ e^{-2 i \sigma\,(R/L)\vartheta}\,F_\sigma(R;\vartheta) + \mathrm{cc}\right]\right\}\,,
\label{x10}
\end{align}
where $\tau_\sigma$ are the polar Pauli matrices. The Coulomb kernel $F_\sigma(R;\vartheta)$
in the polar basis has also been determined in ref.~\cite{Reinhardt:2008ij} by a Fourier expansion
which is strictly periodic in $R$. In the first period $|R| < L$, the series can actually be resummed 
to give the simple result
\begin{align}
  H_C = \frac{g^2}{8}\,\left[ 3 | R | - \frac{R^2}{L} \right]\,\mathbbm{1} \equiv V_C(R)\,\mathbbm{1}\,;
\label{x11} 
\end{align}
for $|R| > L$ it must be continued periodically.
Quite surprisingly, this turns out to be independent of the gauge field $\vartheta$ and can thus be 
pulled out of the gauge-fixed expectation value eq.~(\ref{3}). Inserting again the complete set of 
eigenstates $\langle\vartheta | n \rangle = \Psi_n(\vartheta)$ of $H_{\rm fix}^0$, we arrive at
 \begin{align}
  \langle W \rangle &= \mathrm{tr}\,\langle R | e^{- T (H_{\rm fix} + H_C - E_0)} | R \rangle 
  \stackrel{T\to\infty}{\longrightarrow} \mathrm{tr}\,|\langle 0 | U(R) | 0 \rangle |^2 \cdot e^{- T V_C(R)}\,.
\label{x12}
 \end{align}
The prefactor $Z_W = \tr\,|\langle 0 | U(R) | 0 \rangle |^2$ can be computed explicitly and turns out to 
$T$-independent (and finite). Again, $Z_W$ is the OPE renormalization constant 
for the composite operator $W$ and we obtain the renormalized Wilson loop 
$
\,\langle W_R \rangle = e^{ - T V_C(R)}\,.
$
Since $V_C(R)$ is periodically continued for $|R| > L$, we have to go to the limit $L \gg R$ to 
avoid finite size effects, and the effective string tension becomes
 \begin{align}
  \sigma_{1+1} = \left.\frac{d V_C(R)}{dR}\right|_{R=0} = \frac{3}{8}\,g^2\,.
\label{x14}
 \end{align}
This agrees with eq.~(\ref{x7}) and is also the known expression from the literature. 
It remains to explain why the physical potential $V_C(R)$ is periodic in $R$, but \emph{not} invariant
under $R \to L-R$ as it should be on a compactified $[0,L]\simeq S^1$. A careful analysis 
\cite{Reinhardt:2011fq} along the lines in sec.~\ref{sec:general} reveals that the physical potential
between static quarks is \emph{not} given by any single Wilson loop once the space direction is 
compactified. Instead, all the equivalent loops with $R \to R + m L$ ($m \in \mathbbm{Z}$)
compete and the one giving the minimal potential dominates in the limit $T \to \infty$. Taking this 
prescription into account, the level crossing between $|R\rangle$ and $|R - L\rangle$ at $R=L/2$ 
yields a potential $V(R)$ that is both invariant and symmetric about $R=L/2$.  

\subsection{Yang-Mils Theory in $D=3+1$}
\label{sec:ym31}
For the physically most intersting case of YM theory in $D=3+1$, we start again from 
eq.~(\ref{3}) and introduce the usual Coulomb gauge fixing in the Hamiltonian, combined 
with the resolution of Gau\ss{}' law in the $q\bar{q}$ sector to which the
Wilson state $|\vek{R} \rangle$ belongs,
\begin{align}
\langle W \rangle = \mathrm{tr}\,\langle {\vek{R}_\perp} 
  | e^{- T (H_{\rm YM}^{\rm fix} + H_C[\rho_{\rm dyn} + {\rho}] - E_0)}
  | {\vek{R}_\perp} \rangle 
\label{x15}
\end{align}
In view of the approximations below, it is expedient to perform the unitary transformation 
from section \ref{sec:general} and reshuffle the parallel transporter $U[\vek{A}^\perp](\vek{R})$ 
from the Wilson state into the Hamiltonian. As a result, we have
\begin{align}
\langle W \rangle =
\mathrm{tr} \,\langle 0 | e^{- T (H_{\rm YM}^{\rm fix}\big|_{\Pi \to \Pi + \epsilon} 
  + H_C[\rho_{\rm dyn} + \rho_{\rm ind} + \rho] - E_0)} | 0 \rangle
\equiv \tr\,\langle 0 \,| \, e^{- T (\widetilde{H}_{\rm fix} - E_0)} \,|\,0 \rangle\,,
\label{x16}
\end{align}
where the momentum $\Vek{\Pi}$ in the gauge-fixed vacuum Hamiltonian is shifted by the induced 
electric field eq.~(\ref{6}), and the Coulomb term eq.~(\ref{11})
will now contain an induced charge $\rho_{\rm ind} = [- \vek{A}^\perp, \Vek{\epsilon}^\perp]$, 
in addition to the dynamic charge $\rho_{\rm dyn} = [- \vek{A}^\perp, \Vek{\Pi}^\perp]$ of the gluon 
and the external charge $\rho$ of the static quarks. To proceed, we decompose the complete Hamiltonian 
$\widetilde{H}_{\rm fix}$ from eq.~(\ref{x16}) into three pieces, 
$\widetilde{H}_{\rm fix} = H_{\rm fix}^0 + \widetilde{V}_C + \Delta H$, 
where $H_{\rm fix}^0$ is the usual Coulomb Hamiltonian in the absence of external charges, 
$\widetilde{V}_C$ is the non-Abelian Coulomb interaction between the external charges, 
and $\Delta H$ containing the remaining (indirect) charge interactions.  

In order to evaluate eq.~(\ref{x16}) with this decomposition, we have resort to a sequence of 
approximations, which are not all under good control:

\bigskip
  \begin{tabular}{c@{\,\,}l@{\,\,}l}
    1. & \emph{Jensen's inequality:} &      
          $\langle W \rangle \ge \tr\,\exp\left[ - T \langle 0 | \widetilde{H}^{\rm fix} - 
          E_0 | 0 \rangle \right]  = \mathrm{tr}\,\exp\left[ - T \langle 0 |\, \widetilde{V}_C 
          + \Delta H \,| 0\rangle \right] $
       \\[3mm]
    2. & \emph{Abelization:}         & 
       $[U^\perp(R)\,,T^a] \approx 0$
       \\[3mm]
    3. & \emph{Factorization:}       & 
       $\langle 0 | Q^a\,F^{ab}\,Q'^b | 0 \rangle \approx \langle 0 | Q^a\,Q'^b | 0 \rangle\,
      \langle 0 | F^{ab} | 0 \rangle \equiv \langle 0 | Q^a\,Q'^a | 0 \rangle \,\bar{F}$
  \end{tabular}

%
\bigskip\noindent
Approximation (1.) allows to drop $H_{\rm fix}^0$, while 
(2.) removes terms involving the conjugate momentum $\Vek{\Pi}$ and $\Vek{\epsilon}^\perp$, 
and (3.) simplifies the remaining pieces in $\Delta H$ and, in particular, the Coulomb interaction, 
$\widetilde{V}_C$. 
After a lengthy calculation \cite{Reinhardt:2011fq}, the Wilson loop is thus reduced to a form 
which requires, as only input from the ground state, the gluon propagator 
$\langle 0 | A_i^a(\vek{x})\,A_j^b(\vek{y}) | 0 \rangle = \delta^{ab}\,t_{ij}\,D(\vek{x}-\vek{y})$ 
and the \emph{vev.}~of the Coulomb kernel, $\langle 0 |F^{ab}(\vek{x}-\vek{y}) | 0 \rangle  = 
\delta^{ab}\,\bar{F}(\vek{x}-\vek{y})$:
\begin{align}
\langle \,W\,\rangle \approx& \exp\Bigg\{ - T\,\frac{g^2}{2}\,N_C\int d^3(x,y)\,
\Vek{\epsilon}_a^\perp(\vek{x})\,\Big[ \,N_C^{-1}\,\delta(\vek{x}-\vek{y}) +
\overline{F}(\vek{x}-\vek{y})\,D(\vek{x}-\vek{y})\Big]\,\Vek{\epsilon}_a^\perp(\vek{y})
\Bigg\} \times
\nonumber \\[2mm]
&{} \times\exp\Bigg\{ - T \,C_2\,g^2 \,\Big[ \overline{F}(R) - \overline{F}(0)\,\Big]\,\Bigg\}\,.
\label{x17}
\end{align}
The first term in the exponent is again the self-energy of the Wilson lines which gives rise to a 
renormalization $Z_W$ of the loop operator. The large distance behaviour of the second term is 
dominated by the small momentum form of the correlators, for which we use the relations 
$D(k) \sim [k^2 + M^4 / k^2]^{-\frac{1}{2}}$ and $\bar{F}(k) \sim k^{-4}$ suggested by both 
variational calculations \cite{Epple:2006hv} and recent lattice simulations \cite{Burgio:2012bk}. 
It can then be shown that the second piece in the exponent of eq.~(\ref{x17}) is subdominant 
$(\sim R^{-1})$ at large distances.
The dominant contribution to the Wilson potential is given, in our approximation, by the 
explicit Coulomb interaction, and the Wilson string tension equals the Coulomb string tension, 
$\sigma_W \approx \sigma_C$. 
 
\section{Conclusions}
\label{sec:conclusions}
In this talk, I have presented a recent calculation of the Wilson potential in the Hamiltonian 
approach to Yang-Mills theory. A careful analysis of Gau\ss{}' law and the physical degrees of 
freedom combined with unitary transformations to take care of time evolution reveal that 
the spurious divergences in the Wilson potential can be absorbed in OPE renormalizations of the 
composite loop operator. Using this formalism, the correct result could be derived in cases 
where the exact ground state is known. For the physically most interesting case of Yang-Mills
theory in $D=(3+1)$, only an approximate solution could be obtained in which the Wilson and 
Coulomb string tension equal (while $\sigma_W/\sigma_C \simeq 0.3\,\ldots\,0.5$ in 
lattice simulations).
The main reason for this discrepancy is most likely the approximate \emph{factorization} 
used in the calculation, which  neglects the screening of external charges by gluons. 
One-loop calculations of this contribution indeed indicate a reduction of $\sigma_W$, and 
further investigations of this issue are in preparation.


\end{document}